# Self-assembly of micro-machining systems powered by Janus micro-motors

C. Maggi[1], J. Simmchen[2], F. Saglimbeni[1], J Katuri[2,3], M. Dipalo[5], F. De Angelis[5], S. Sánchez [2,3,4] and R. Di Leonardo[1,6]

[1] Dipartimento di Fisica, Università di Roma "Sapienza", I-00185, Roma, Italy

[2] Max-Planck Institute for Intelligent Systems, Heisenbergstr. 3, 70569 Stuttgart, Germany

[3] Institute for Bioengineering of Catalonia (IBEC) Baldiri i Reixac, 10-12 08028 Barcelona (Spain)

[4] Institucio Catalana de Recerca i Estudis Avancats (ICREA), Passeig Lluis Companys 23 08010 Barcelona (Spain)

[5] Istituto Italiano di Tecnologia, I-16163, Genova, Italy and

[6] NANOTEC-CNR, Institute of Nanotechnology, Soft and Living Matter Laboratory, Piazzale A. Moro 2, I-00185, Roma, Italy

Integration of active matter in larger micro-devices can provide an embedded source of propulsion and lead to self-actuated micromachining systems that do not rely on any external power or control apparatus. Here we demonstrate that Janus colloids can self-assemble around micro-fabricated rotors in reproducible configurations with a high degree of spatial and orientational order. The final configuration maximizes the torque applied on the rotor leading to a unidirectional and steady rotating motion. We discuss how the interplay between geometry and dynamical behavior consistently leads to the self-assembly of autonomous micromotors starting from randomly distributed building blocks.

## 1. Introduction

The manipulation, transport and assembly of micro-objects is of paramount importance for micro-engineering and biological applications. Among the earliest approaches were microelectromechanical systems (MEMS) based microgrippers.[1, 2] They represent a very versatile tool that might be used for handling fragile objects from millimeter to micron size.[3, 4]

Non-contact manipulation can be achieved by electrostatic actuators evolved from tip like actuators [5] to cilia inspired structures.[6] The use of magnetic fields represents an elegant solution for the contact-less manipulation of magnetic microobjects.[7-9] . For example Tottori et al. demonstrated helical micromachines with microholders that used mechanical contact to transport passive particles.[9] Alternatively, holographic optical tweezers can be used for precise, three dimensional micromanipulation of multiple dielectric objects.[10, 11] Those methods accurately control grasping forces in order to avoid any damage to small-sized delicate structures, yet they require in general very large, complex devices and conditions that are usually met only in research laboratories. In contrast, self-propelled micromotors can be employed for micromanipulation and transport avoiding the use of any external field. Different catalytic rod- and tube-shaped micromotors were shown to be useful for cargo transport,[9, 12-22] while Janus particles have been shown to be capable of moving particles larger than their size.[23-25] Similarly, it has been shown that living kinds of active matter, such as bacteria, algae or sperm cell suspensions, can be used to propel larger structures [26-38] or delivery colloidal particles onto target sites.[30] Suspensions of bacteria or synthetic micro-swimmers show fascinating collective behaviors which, depending on concentration of bacteria, might give rise to hydrodynamic turbulence (swirling state). Kaiser et al. found that both polar ordering and swirl shielding inside a wedge structure can yield an optimal transport velocity which becomes even bidirectional at high concentrations.[39] All phenomena mentioned above take place only at relatively high bacterial concentrations and always involve a partial rectification that is inevitably accompanied by a high degree of randomness and low reproducibility. Here we demonstrate that a small and well defined number of artificial micromotors can self-organize in highly ordered configurations around a larger passive microobject and propel it in a steady unidirectional motion. The highly deterministic character of the proposed self-assembly strategy is based on a previously reported dynamical behavior of Janus particles colliding with solid obstacles [40] and it is similar to a dynamic assembly behavior observed between (catalytic) particles.[41] When approaching walls particles orient their symmetry axis parallel to the wall surface because of the interaction of the self-generated solute gradient with both the substrate and the wall. By choosing a convenient relationship between the lengths of the sides of a microgear, we force the active particles to dock on the gears in a well-defined position and orientation. In this manuscript we study the effect of number and orientation of the particles on the motion of the microgears. We envision that the use of Janus particles as self-assembling actuators for larger microobjects, could pave the way for more applicable micromachines.[40]

## 2. Results and Discussion

The active component is made of platinum-coated silica Janus particles (5 μm diameter) that can self-propel in a mixture of deionized water and hydrogen peroxide. The passive component is a micro fabricated gear having six asymmetric teeth with an external radius of 8 μm (gray area figure 1). The two components, initially in a deionized water solution, are mixed on a plasma-cleaned glass coverslip where they sediment displaying rotational and translational Brownian motion restricted to the proximity of the surface. The typical final area concentrations range from 1.1 to $9.7 \times 10^{-3}$ μm$^{-2}$ for Janus particles and are about $5.5 \times 10^{-5}$ μm$^{-2}$ for micro-gears. When hydrogen peroxide ($H_2O_2$, 5% concentration) is added to the sample we immediately observe active motion of the Janus particles. The sample is observed by bright-field microscopy recording digital images at a frame rate of 50 fps (see Experimental Section). Acquisition of digital frames is always performed few seconds after the addition of fresh $H_2O_2$ to the sample in order to maintain the activity of the Janus particles approximately constant for all measurements. Few seconds after $H_2O_2$ addition, one or more Janus particles collide with each microgear. The active particles align their propelling direction along the side of the gear and, depending on their incident angle with the edge normal, either leave the structure or slide along the edge until they get stuck on a corner. Two distinct final configurations are possible (see **Figure 1**).

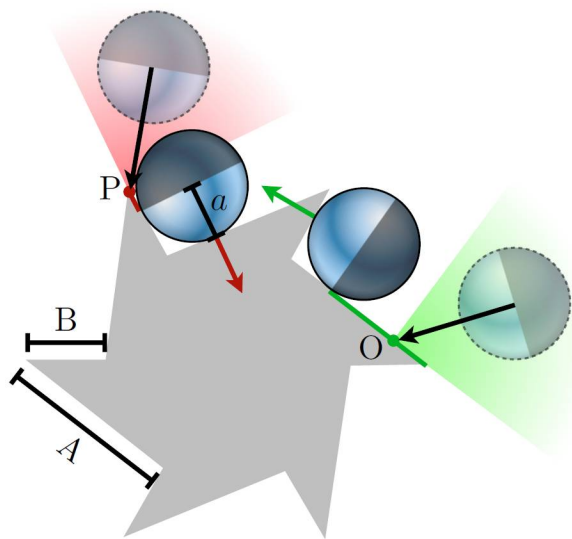

**Figure 1.** Schematic representation of the two possible docking events leading to a Janus particle stably propelling the microgear.

A particle collides in O coming from the green shaded region, aligns and then eventually docks propelling the structure in Figure 1 in an anticlockwise direction. The probability of events of this type is proportional to the length A-a of the green segment representing the locus of possible impact points O. A second docked state occurs when a particle collides in P coming from the red region and gets stuck pushing the structure in the opposite direction and with a lower torque. Analogously the probability of these events is proportional to B-a. Therefore the majority of fully occupied structures will have all particles pushing in the same anticlockwise direction, while the fraction of "wrong" assemblies where for example one particle points in the opposite direction is only (B-a)/(A-a) with the optimal situation being B ~ a. In our case B= 3 µm, A= 6.7 µm so that only one structure among 10 is expected to form with a misoriented particle. The alignment observed here is consistent with previous experiments on the alignment of Janus particles along extended linear obstacles of various thickness, and it has been ascribed to the influence of the boundary onto the phoretic slip flow generated by the particle.[40] **Figure 2** reports a typical docking event for a 5 µm Janus particle (see also **Supplementary video S1**) which is composed of three main stages: in Figure 2(a) the particle moves freely pointing towards a gear, Figure 2(b) the particle touches the long edge and aligns along it, finally in Figure 2(c) the particle slides along the long edge until it arrives at the short edge and starts pushing with maximal applied torque.

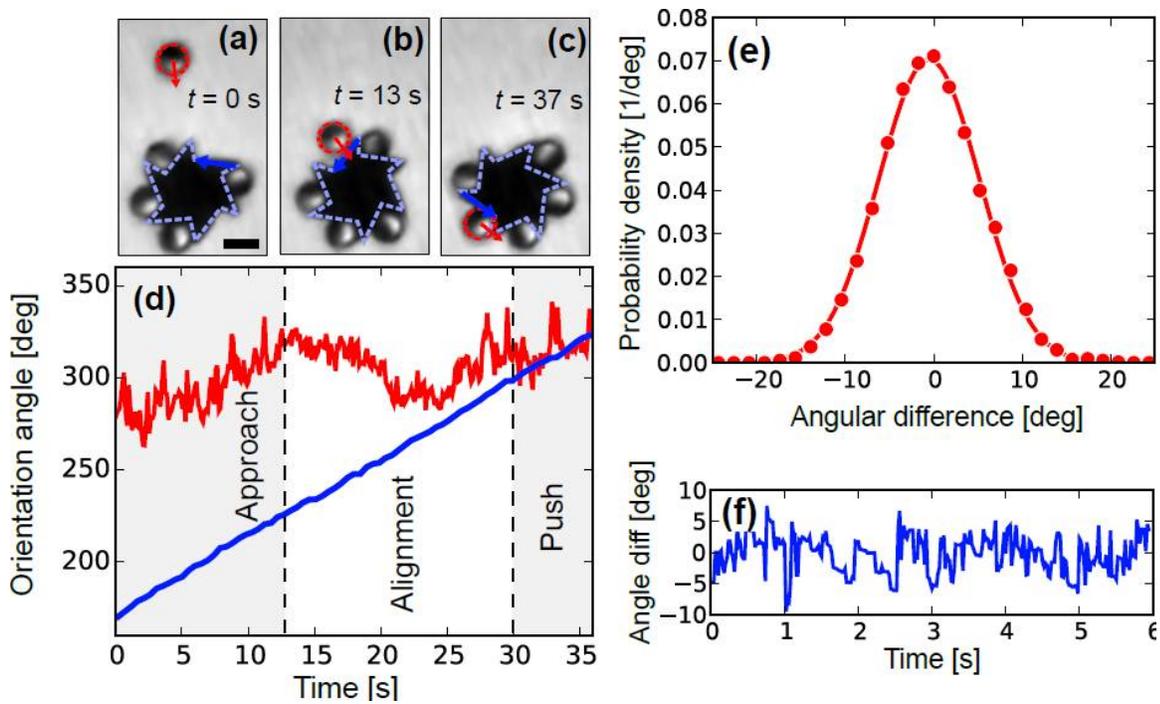

**Figure 2.** (a) Janus particle approaching the gear. Dashed lines represent the tracked shapes, arrows indicate the instantaneous orientation of particle propulsion and of the colliding long edge (scale bar is 5 μm). (b) Collision followed by alignment. (c) The Janus particle docks on the gear with an orientation resulting in a maximal applied torque. (d) Temporal evolution of the orientation angle for the gear (blue line) and of the Janus particle (red line), during the approach, alignment and pushing stages. (e) Probability distribution of the angle formed by a docked Janus particle and the corresponding long edge. (f) Typical uctuations of the angle formed by the orientation vector of the Janus particle and the corresponding long-tooth edge as a function of time.

This is not the case when the particles and structure size are not matched. For example, using particles of smaller size (2 μm diameter) leads to frequent misalignment (see **Figure 3**).

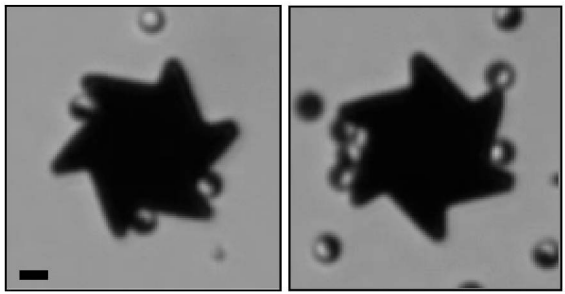

**Figure 3.** Two snapshots of the microgear in a suspension of 2 μm Janus particles (scale bar is 2 μm). The particles are often found misaligned with respect to the gear's edge.

Moreover, since rotational diffusion decreases as the inverse cubic size, smaller particles reorient very rapidly resulting in a shorter lifetime of docked states (see **Supplementary video S2**). In order to better investigate the self-assembly process we only focus on particles having the optimal size of 5 μm. A shape-recognition algorithm is used to track the gear and the Janus particles. We define particle and gear edge orientation as indicated by arrows in Figure 2(a-c). The angle formed by these two vectors with a common horizontal reference is shown as full lines in Figure 2(d). Before collision the Janus particle only shows moderate fluctuations of its orientation (approach stage) while the gear rotates at approximately constant speed under the action of four well-aligned particles. After collision we observe reorientation of the particle (alignment stage). Once the alignment is complete and the particle reaches the short blocking edge the two orientation angles merge on the same curve and the assembly moves as a rigid body (push stage). Once the particle is trapped in the corner of the gear its orientation is stably locked around the orientation of the long edge of the gear showing only small fluctuations. This has been quantified by tracking for at least 5 s the orientation of about 30 different Janus particles fitting in the corners of about 10 different microgears and the orientation relative to the nearest long edge. In Figure 2(e) we plot the histogram of the angular difference between the orientation angle of the Janus particle and the orientation angle of its nearest long edge (see arrows in Figure 2(a-c)). The probability distribution of this angular difference is well fitted by a Gaussian (full line in Figure 2(e)) resulting in average angular fluctuations of 5.7 +/- 0.8 degrees. The amplitude of fluctuations is very similar for all Janus particles tracked, a sample of these fluctuations is shown in Figure 2(f) where we report the angular difference for one single particle for a time span of 6 s. When a Janus particle collides, aligns and docks to the gear we usually see a clear increase of the rotational speed. Occasionally we also observe that a stuck Janus particle reorients, because of Brownian fluctuations, and leaves the structure. These events are instead accompanied by a decrease in rotational speed of the gear. In both cases the change of the rotational speed is approximately proportional to the free propulsion speed of the Janus particle far from the structure. We measure the change of rotational speed and the speed of the incoming/leaving Janus for six different attaching/detaching events happening at low Janus particles density (when at most three Janus particles are in contact with the gear). These measurements are obtained by tracking gears and particles for at least 3 s. Note that we start measuring the change of rotational speed of the gear, caused by an attaching Janus particle, only when the latter becomes well aligned with long-tooth edge. In this way we can ignore the effect of initial orientation of the incoming particle on the gear speed. The results are shown in **Figure 4** where we plot the change of rotational speed of the microgear, after one single event, as a function of the speed of the particle.

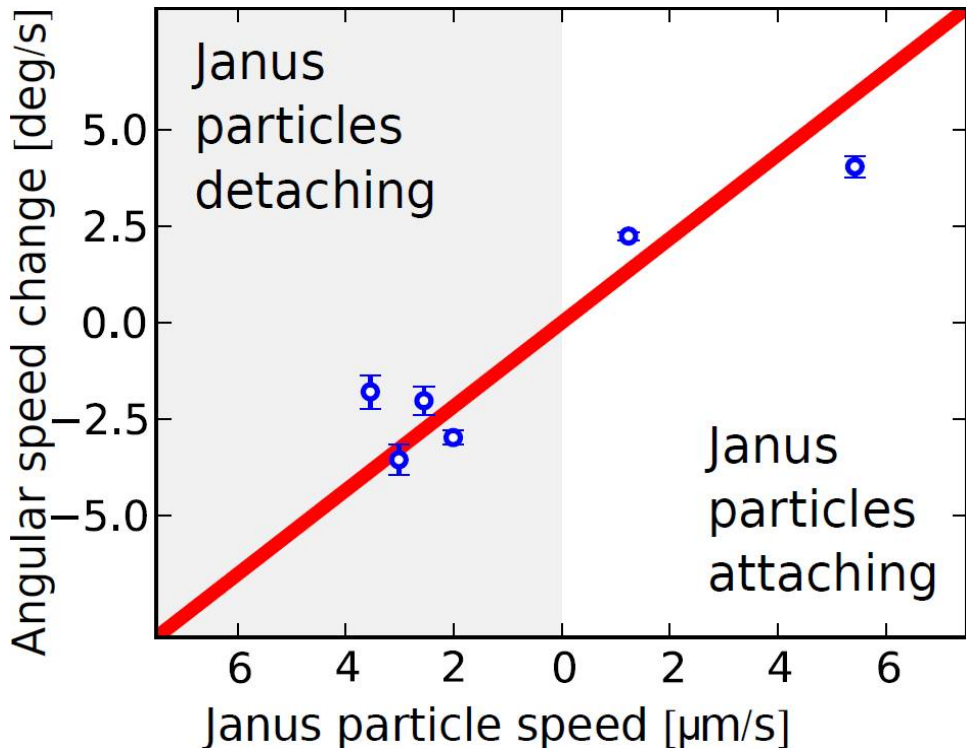

**Figure 4.** (a) Change in angular speed (circles) of the microgear when one single Janus particle attaches (right) or detaches (left) as a function of the particle speed as measured before attachment or after detachment. The full line is a linear fit passing through zero.

The angular coefficient of the linear t shown in Figure 4 (full line) results to be α= 1:1 ± 0:1 deg/μm. A rough estimate of this factor can be obtained if we assume that a pushing Janus particle exerts a force of order f = 6πηav where v is the particle's speed and η the viscosity of the solvent. This force will generate a torque T = fr, with r the radius of the microgear. By approximating the rotational viscous drag of the Janus-gear structure as Γ~8ηπr³ we obtain the contribution ΔΩ of one single particle to the gear rotational speed as ΔΩ = T/Γ= αv, where α= (3/4)a/r² ~ 2.5 deg/μm which is of the same order of magnitude of the measured value. Upon increasing the Janus particles density the probability of finding more teeth occupied on each single gear grows significantly. By increasing the overall number of Janus particles per unit surface of a about a factor ten (from 0.022 ± 0.06 to 0.19 ± 0.02 area fraction) we observe that the number of particles fitting in the microgear's teeth increases gradually from 1 to 6 (maximum occupancy, see **Supplementary video S3**). This is shown in **Figure 5** (a),(b) and (c) where we show typical configuration at low, intermediate and high particles densities respectively.

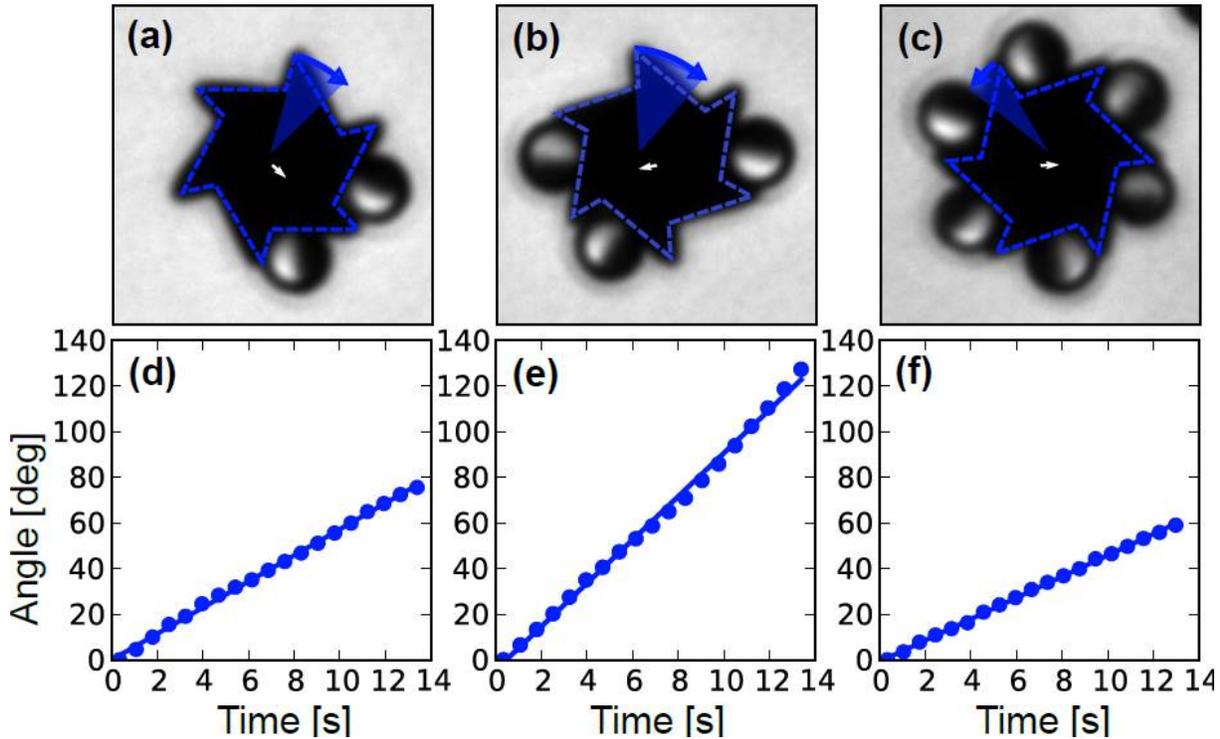

**Figure 5.** (a) One microgear pushed by two Janus particles. The dashed line represents the results of the tracking. The shaded area represents the angle spanned in a time interval of 4 s. The small white arrow represents the displacement of the center of mass in a time span of 14 s. (b) A self-assembled configuration with three occupied sites. (c) A fully formed rotating structure.(d), (e), and (f) Time evolution of the cumulative rotation angle (circles) for the gears shown in (a),(b) and (c) respectively. The lines represent linear fits of the data.

Upon increasing the number of Janus particles in contact with the gear from 1 to 2 to 3 we observe a reproducible increase of the angular speed of the gear as seen by comparing the angle spanned by the gear with 2 and 3 particles in 14 s (shown in Figure 5(a) and (b) respectively). We also observe that the center of the gear shows always a displacement limited to few microns in the measurement time (small arrows in Figure 5(a),(b) and (c)). Upon further increasing the particle density, for a number of occupied sites going from 4 to 6 Janus particles, we observe instead a decrease of rotation rates. This non monotonic behavior of rotational speeds with the number n of occupied sites is already visible in Figure 5 for n = 2, 3, 6 and summarized in **Figure 6** for all values of n.

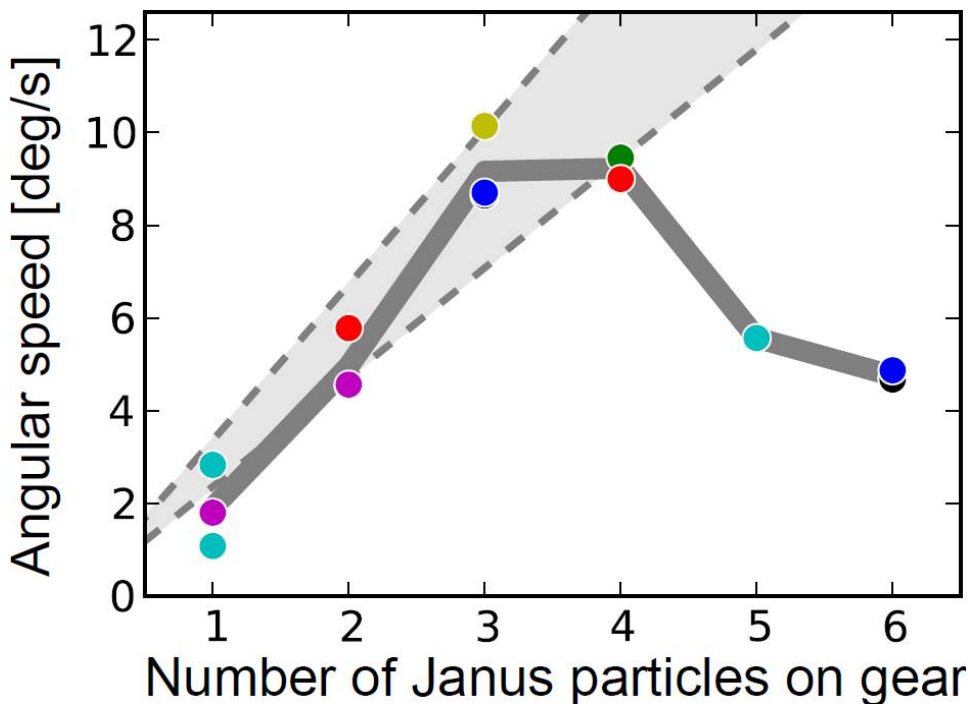

**Figure 6.** Non monotonic behavior of the gears angular speed with the number of self-assembled Janus propellers. Each circle represents a different gear. The full line represents an average over all the gears having the same number of Janus particles in contact. The shaded area is the predicted angular speed from single-event observation (Figure 4, see text).

These last are obtained by tracking the orientation angles of about 10 different microgears for at least 5 s. By averaging the measurement of rotational speed obtained for the same number of particles in contact (thick line in Figure 6) we obtain a clear trend confirming the non-monotonic behavior of gear rotational speed described above (see Figure 5). If the number of Janus particles in contact with the gear is limited to a maximum of 3 the rotational speed increases systematically with the number of particles in contact. The magnitude of this increase is also consistent with the estimate of the contribution given by each particle to the rotational speed shown in Fig 4. By using the measured value of $\alpha = 1.1 \pm 0.1$ deg/µm and the measured average speed of all Janus particles tracked $v = 2.6 \pm 0.4$ µm/s, we expect the angular speed of the gear to grow linearly with the number n of particles in contact as $\Omega = \beta n$ where $\beta = \alpha v = 2.8 \pm 0.5$ deg/s which is shown as a shaded area in Figure 6, which captures the trend of the data up to n = 3. Differently for n > 4 the $\Omega$ stops increasing and then decreases. At a fixed fuel concentration we would expect that the rotational speed could never exceed the value $v = \Omega r$ for which the free propulsion speed of the particle equals the linear speed of the rotor's edge. Additionally the total drag of the rotating body increases with n so that a slow saturation to the maximum speed would be expected. The observed speed decreases at larger n values should most probably be attributed to a progressive reduction of fuel concentration due to both the local and global increase in the number density of Janus particles depleting hydrogen peroxide.

## 3. Conclusion

We have demonstrated the self-organization of self-propelled Janus particles to power passive larger engineered objects, i.e. micro-gears. The main steps of this self-assembly process involve collision with the gear and subsequent alignment and docking. We have shown that an appropriate choice of the gear geometry leads to reproducible final configurations where each single particle contributes maximally to the propulsion of the larger object. The propulsion speed shows a non-trivial behavior with the total number of gathered propellers. The self-assembly process relies on the peculiar behavior of self-propelling particles in contact with confining structures. The use of more sophisticated microfluidic environments allows to tune the concentration of peroxide in Janus particle systems [42] and may offer the possibility to trigger the self-assembly and control the rotational speed of our micromotors. Understanding and exploiting these complex mechanisms may provide novel strategies to design autonomous micro-machines for lab on chip applications.

## 4. Experimental section

*Janus particles synthesis:* Janus particles were obtained by drop casting of a suspension of spherical silica colloids (5 µm diameter, Sigma Aldrich) on an oxygen-plasma cleaned glass slide followed by slow evaporation of the solvent and placed in an electron beam system. High vacuum was applied and subsequently a monolayer of 2 nm Ti was evaporated to guarantee good adhesion of the 7 nm Pt for catalytic properties. To release particles from the glass slides into DI water, short ultrasound pulses were sufficient.

*Substrate treatment:* To obtain clean and hydrophilic substrates glass slides were cleaned with alcohol, dried and treated during 5 min in a Diener oxygen plasma machine. Substrates were used immediately after this procedure.

*Gears microfabrication:* The fabrication starts with the spin-coating and baking (200°C) of LOR3B/SU-8 (200 nm/4 µm on a silicon wafer). An amorphous carbon film (100 nm) is then deposited on the SU-8 by sputtering. Subsequently, LOR3B/S1813 (200 nm/1.5 µm) are spin-coated on the carbon layer, and laser lithography is performed to obtain the negative pattern of the microgears in the LOR3B/S1813 bilayer. After the evaporation of 100 nm chrome, the LOR3B/S1813 bilayer is removed by N-methyl-2-pyrrolidone, leaving chrome microgears on the carbon layer. The SU-8/carbon layer is then etched by Reactive Ion Etching (RIE) in oxygen plasma; the chrome microgears act as etching mask and transfer their shape onto the SU-8; the etching process is anisotropic and produces SU-8 vertical walls. A picture of the finalized SU-8 microgear can be found Figure 1 of Ref. [43] after chrome removal in chrome-etch solution. The microgears are finally released from the wafer by PG-remover, which dissolves the sacrificial LOR3B layer at the bottom of the microgears (see Ref. [43] for more details about the microfabrication process).

*Optical setup:* Bright-field microscopy is performed by using an inverted optical microscope (Nikon TE-2000U) equipped with a 20x (NA=0.5) objective. Images are recorded with a high-sensitivity CMOS camera (Hamamatsu Orca Flash 2.8). In each measurement we record videos at least 3 s long (150 frames) up to a maximum of about 30 s (1500 frames).

**Acknowledgements**

The research leading to these results has received funding from the European Research Council under the European Union's Seventh Framework Programme (FP7/ 2007-2013)/ERC Grant Agreement No. 307940 and No. 311529, LT-NRBS. S.S., J.K. and J.S. thank the DFG for financial support (SA2525/1-1).